# Influence of process conditions and pore morphology on the closure rate of pores in hot rolling of steel

C. Liebsch[1*], X. Li[1], J. Lohmar[1] and G. Hirt[1]

[1]Institute of Metal Forming, RWTH Aachen University,
Intzestraße 10, 52072 Aachen, Germany
*Corresponding Author, conrad.liebsch@ibf.rwth-aachen.de

**Abstract**

Steel sheets are manufactured from slabs produced in continuous casting, which inevitably results in a porous initial microstructure. These pores are nuclei for ductile damage and need to be closed during rolling in regard of the strict performance requirements of todays advanced high strength steels. Due to the beneficial shape factor of the roll gap, recrystallization and high diffusion rates pore closure and elimination occurs primarily during hot rolling. In this paper, relevant influencing factors on pore closure are investigated based on representative volume elements (RVE). First, the load regime of typical multi-pass hot rolling schedules is calculated via macro scale FEM and then applied to micro scale RVEs. Within the RVE, shape changes of a given pore are tracked and the closure ratio is calculated. This paper elucidates the effect of different roll diameters, absolute reductions per pass, pore shape and pore orientation on the closure rate. The results show that the majority of pores can be closed during the initial hot rolling stages. However, pore closure can be accelerated through suitable process parameters. Furthermore, shape and orientation of pores should be taken into account, if a precise knowledge of the total height reduction, leading to complete pore closure, is desired.

## 1. Introduction and State of the Art

A certain amount of initial porosity is inevitably introduced into steel slabs by continuous or ingot casting prior to hot working. This porosity is a prime nucleation site for ductile damage and finally failure during further rolling into sheets. As almost all produced steel is hot rolled after casting, it is desirable to remove as much porosity as possible already at this stage. This is promoted due to favorable conditions in terms of pore closure during hot rolling i.e. high temperatures and mostly compressive hydrostatic stresses [1]. Furthermore, recovery, recrystallization and high diffusion rates enable the interface to fully dissolve and thus heal the microstructure as shown by Park and Yang [2].

The load regime i.e. the local stress components influence the closure behavior of pores during rolling. The load regime in turn mainly depends on the shape factor, influenced by roll radius and pass reduction as well as the considered position within the roll stock [1]. Already in 1987 Tanaka et al. [3] proposed a criterion to determine the closure probability of spherical pores in dependence of the stress triaxiality, which can be seen as the hydrostatic stress normalized by the equivalent stress. Recently, Liebsch et al. [4] used this criterion to investigate the pore closure during hot rolling of a dual phase steel.

In 2015 Saby et al. [5] put forward a mesomechanical model that also recognized the importance of the pore shape and orientation for the closure behavior of pores during forming. However, this model requires several parameters that are not readily available since they have to be determined from various simulations of proportionally loaded RVEs. A more precise technique to access the influence of the process conditions immediately as well as the shape and the orientation of the pores is a direct transfer of the local non-proportional boundary conditions (BC) to RVEs containing explicit pores. Saby et al. [6] suggested that this technique is applicable if the porosity in the material is small enough not to deteriorate its macroscopic properties.

The aim of this paper is to model hot rolling of a dual phase steel and to investigate the resulting pore closure behavior using RVEs. More specifically, it aims to help improve the robustness of pore closure prediction in hot rolling. At best, all pores should be closed already during hot rolling. To enable this, pore shapes as well as orientations that are hardest to close need to be known. When simulating the closure behavior these values can be estimated by comparing the results of a spherical pore within the center of the slab and two ellipsoidal pores (prolate; oblate) both orientated in all principle directions. Additionally, investigating the closure of a spherical pore under different rolling conditions provides a direct comparison of the impact that pore morphology and process variables have on pore closure in the center of the slab. In the future, this investigation will be extended to also consider the influence of different load regimes close to the slab surface.

## 2. Methods and Procedure

To model pore closure in hot rolling two distinct types of FE models are required. The first model is a macroscopic process model that provides the BCs in the roll gap during rolling. When disregarding spread material flows primarily in rolling direction (RD) making a 2D plane strain FE model sufficient to describe the process. The BCs determined in the center of the slab are subsequently fed into a second 3D FE model that explicitly contains a pore with a defined shape. This model is referred to as representative with respect to the behavior of the pore it models and is thus called an RVE. A more detailed description of the coupling procedure between macroscopic process model and RVE is omitted here due to space constrains but is given by Wei et al. in [7].

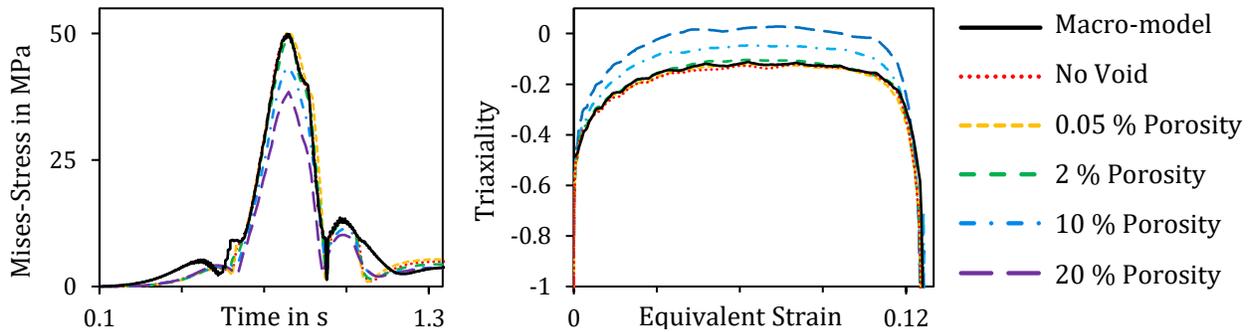

*Figure 1. Equivalent stress and triaxiality of the macroscopic model and the RVEs*

To model pore closure the RVE needs to violate volume conservation, which is typically required in metal forming simulations. This is achieved by transferring a mixed set of BCs from the macroscopic model to the RVE. Displacement is applied via the displacement gradient of the considered element and is used in normal (ND) and transverse direction (TD). In contrast, stress BCs in RD provide the necessary degree of freedom.

To validate that the used RVE satisfies the requirements suggested by Saby et al. [6] and mentioned earlier the Mises-stress and triaxiality resulting from the macro-model are compared to the values obtained from RVEs with different porosities, each represented by a single void in its center. The results in Figure 1 show that differences between the macroscopic model and the RVEs should be negligible if the pore volume fraction is < 2 %. Based on experimental data the following study assumes a porosity of 0.05 %.

## 3. Results and Discussion

To investigate the pore closure behavior hot rolling of a DP800 dual phase steel slab from 100 to 60 mm thickness at an initial temperature of 1200 °C is simulated. The work rolls are 410 mm in diameter and the reduction in every pass is 10 mm. These conditions were chosen as they enable a subsequent validation using the laboratory rolling mill at the IBF.

Figure 2 summarizes the closure curves i.e. the evolution of the pore volume with strain obtained under the aforementioned conditions for two ellipsoidal pores, one prolate (left) and one oblate (right) in comparison to a spherical pore with identical volume. The Figure shows the results when orienting these pores in the three principle directions.

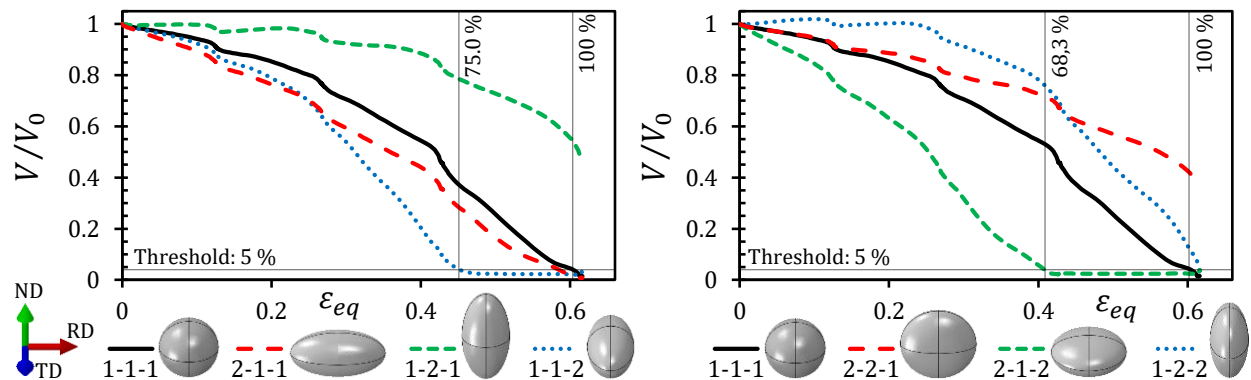

*Figure 2. Closure curves in dependence of pore shape and orientation*

Figure 2 illustrates that different shapes and orientations have a great influence on the necessary strain to full pore closure. Non-spherical pores with at least one elongated axis oriented in ND are particularly hard to close. In contrast, those with their elongated axes oriented in RD or TD are closed more easily than a spherical pore. Both results are in line with expectation as the axis dominating the pore closure behavior is always oriented in normal direction. Furthermore, due to a more advantageous shape evolution throughout the process pores with an RD/TD ratio of 2/1 (2-1-1; 2-2-1) show a linear-like closure behavior while pores with an RD/TD ratio of 1/2 (1-1-2; 1-2-2) close rather exponentially.

To put the impact of pore shape and orientation into perspective Figure 3 shows the influence of roll diameter (left) and absolute reduction per pass (right) on the closure behavior of a spherical pore. The direct comparison reveals that the influence of these process parameters on the strain to closure is lower than a change in pore shape or orientation. To quantify this insight the strain to closure normalized by the spherical pore's strain to 95 % closure was calculated for all closure curves. Depending on the orientation, a prolate pore can be closed after 75.0 % and an oblate pore after 68.3 %. For a different orientation, both pores can remain half-open. In contrast, varying roll radius or absolute reduction per pass can only achieve 91.3 % and 88.4 % relative strain to closure respectively and thus displays a much tighter range of closure rates.

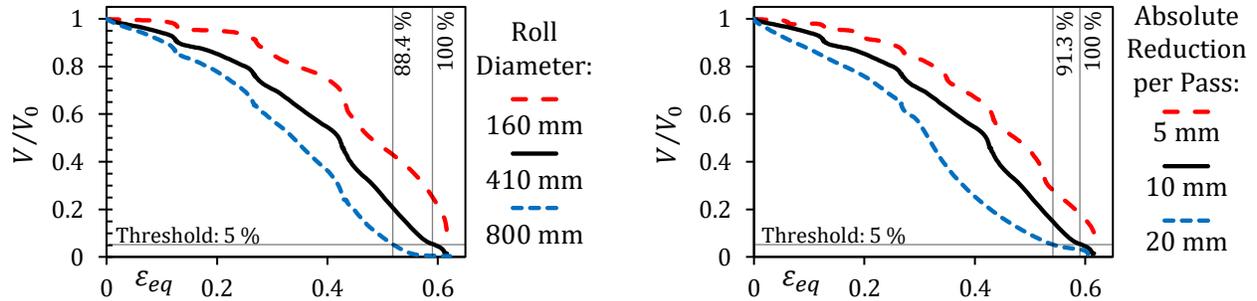

Figure 3. Closure curves for roll diameter (left) and absolute reduction per pass (right)

## 4. Summary and Conclusion

In this paper, the use of RVEs to predict pore closure in hot rolling of a dual phase steel was demonstrated. It was shown that the shape of a pore and its orientation within the center of a steel slab has a greater influence on its closure probability during rolling than the process conditions used. Whether this is also true for pores close to the slab's surface will be the focus of future research. However, pore closure can be accelerated through suitable process parameters. Ellipsoidal pores with an elongated axis orientated in normal direction are particularly hard to close during rolling. Thus, the behavior of such pores in dependence of the process conditions should be studied in detail when aiming for a safe metric that predicts complete pore closure.

## 5. Acknowledgements

The authors gratefully acknowledge the funding by the Deutsche Forschungs-gemeinschaft (DFG, German Research Foundation) for the project A04 within the TRR 188 – Project number 278868966 on "Damage Controlled Forming Processes".

## 6. References

[1] Llanos JM, Santisteban V, Demurger J, Kieber B, Forrestier R, et al. (2008) Improvement of central soundness in long products from a through process control of solidification and reheating and rolling parameters. Tech. Rep. European Commission - Research Fund for Coal and Steel.
[2] Park CY, Yang DY (1996) A study of void crushing in large forgings I: Bonding mechanism and estimation model for bonding efficiency. JMPT 1996;57
[3] Tanaka M, Ono S, Tsuneno M, Iwadate T (1987) An Analysis of Void Crushing during Flat Die Free Forging. Paper presented at 2nd Int. Conf. on Techn. of Plast., Stuttgart, 24-28 August 1987
[4] Liebsch C, Hirt G (2018) Numerical investigation on damage evolution and void closure in hot flat rolling. Paper presented at 11th Forming Technology Forum, Zurich, 2-3 July 2018
[5] Saby M, Bouchard P-O, Bernacki M (2015) A geometry-dependent model for void closure in hot metal forming. Finite Element in Analysis and Design 2015
[6] Saby M, Bernacki M, Roux E, Bouchard P-O (2013) Three-dimensional analysis of real void closure at the meso-scale during hot metal forming processes. Comp. Mat. Sc. 2013
[7] Wei X, Hojda S, Dierdorf J, Lohmar J, Hirt G (2016) Crystal plasticity finite element analysis of texture evolution during cold rolling of a non-oriented electrical steel. Paper presented on 10th Int. Roll. Conf., Graz, 6-9 June 2016